# Laser-design methodology for extending an available wavelength range of dual-wavelength injection-seeded laser

Ayaka Hashimoto,[1] Tetsuto Watanabe,[1] Sota Kobayashi,[1] Megumi Otose,[1] Ken'ichi Nakagawa,[1,2] AND Masayuki Katsuragawa[1,2,3,4,*]

[1]*Department of Engineering Science, University of Electro-Communications, 1-5-1, Chofugaoka, Chofu, Tokyo 182-8585, Japan*
[2]*Institute for Laser Science, University of Electro-Communications, 1-5-1, Chofugaoka, Chofu, Tokyo 182-8585, Japan*
[3]*Institute for Advanced Science, University of Electro-Communications, 1-5-1, Chofugaoka, Chofu, Tokyo 182-8585, Japan*
[4]*National Institute of Polar Research, 10-3, Midoricho, Tachikawa, Tokyo 182-8585, Japan*
[*]*katsuragawa@uec.ac.jp*

**Abstract:** This paper describes a method for expanding the applicable wavelength range of dual-wavelength oscillation injection-locked lasers. Special focus is placed on the laser design that enables simultaneous oscillations at both the gain center and gain tail, requiring the most stringent requirements. We show that a sufficient performance can be achieved by introducing a wavelength-dependence property onto the output coupling mirror, without giving any losses in the laser resonator. We also show how we can determine such wavelength-dependent property on the basis of experimental measurements. Using the designed output-coupler, in practices, we demonstrate simultaneous oscillations with an equal output power, across a variety of wavelength combinations of near the gain center and around the gain tail in which the wavelength separation amounts 60 nm at the most.

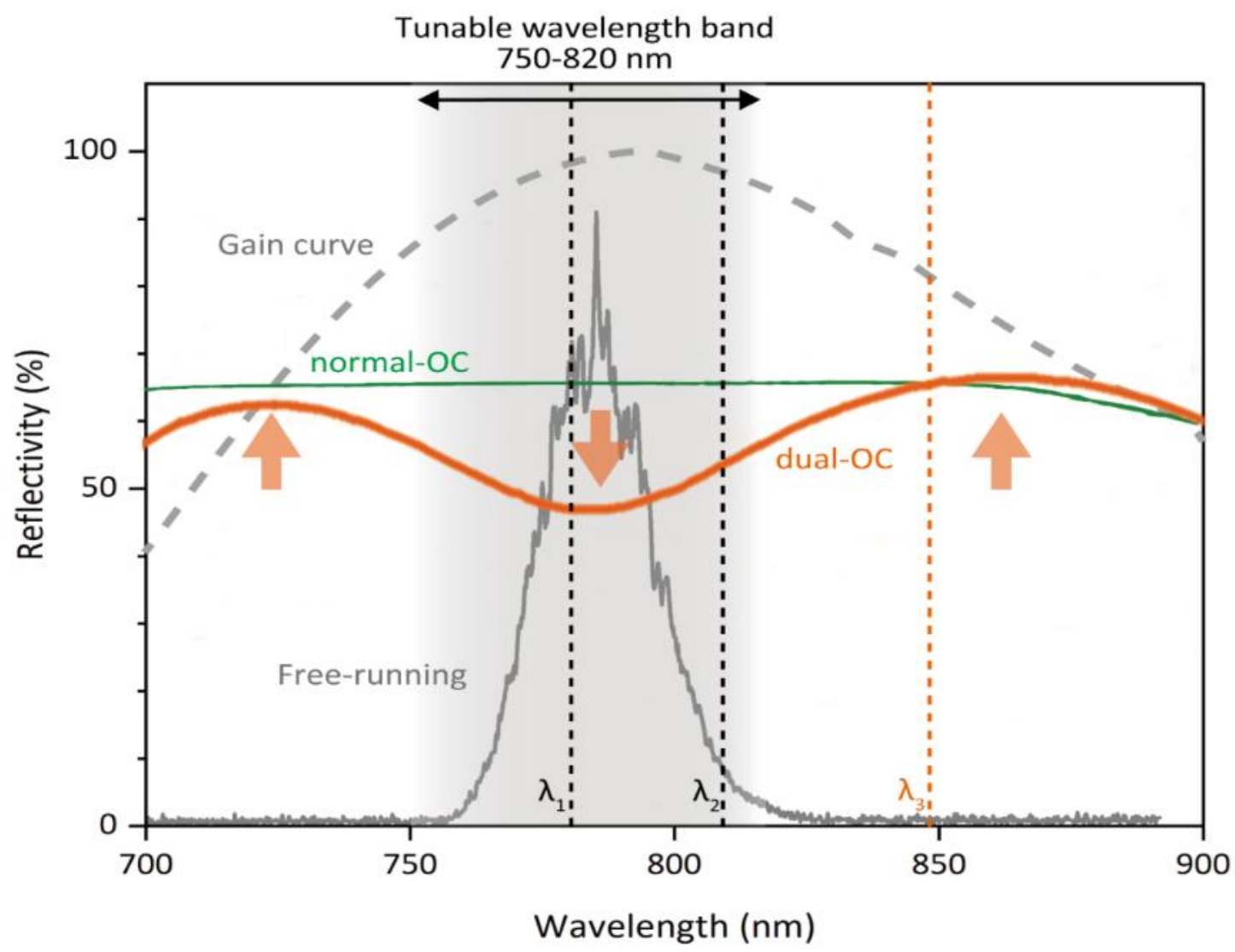

**Fig. 1. Fundamental properties of a Ti: sapphire laser and the conceptual idea for extending an available wavelength bandwidth of the dual-wavelength injection-seeded Ti: sapphire laser.** Gray dotted line, the gain curve of Ti: sapphire laser crystal. Gray spectrum, free-running oscillation spectrum by employing a standard output coupler (normal-OC, green curve) having a property of constant reflectivity of 66.5% over 750 - 850 nm. Orange curve, an expected reflectivity property of the designed output coupler (dual-OC) for extending an available wavelength bandwidth of dual-wavelength oscillation.

Injection-locked lasers are an exceptionally versatile laser system that achieves both high frequency-purity and high output-power, simultaneously [1] [2] [3] [4] [5]. While their fundamental operation is based on single-wavelength oscillation, the conceptual idea can be extended to multi-wavelength oscillations. Indeed, such conceptual ideas have been successfully implemented in both nanosecond pulsed [6] [7] [8] [9] and continuous-wave lasers [10]. In studies involving nonlinear optical processes, laser radiations having different wavelengths are generally involved. Multi-wavelength injection-locked lasers, thus, serve as particularly useful tools in such applications [11] [12].

The combination of wavelengths can be optimally selected depending on specific applications. However, the available wavelength range allowed for a unique laser system is not necessarily broad. Figure 1 illustrates the case of applying dual-wavelength operation to a titanium-doped sapphire (Ti: sapphire) laser which has a very wide gain bandwidth in the near-infrared region (670 – 1000 nm) [13]. In a standard laser configuration using a normal output-coupling mirror with a constant reflectivity-characteristics in terms of wavelength, the available wavelength range is typically limited to approximately 750 – 820 nm. For example, dual-wavelength operation is achieved for the wavelength combination of $\lambda_1$ (790 nm) and $\lambda_2$ (820 nm), but if we select a wavelength combination of $\lambda_1$ and a wavelength further away from the gain peak, such as $\lambda_3$ (840 nm), the laser output is no more pulled into the wavelength, $\lambda_3$ [14], thereby, the dual-wavelength oscillation is not achieved.

Expanding the available wavelength range operated with the multi-wavelength injection-seeded oscillation may significantly broaden the potential applications, thus a valuable challenge in the laser technology. Here, we describe the methodology of expanding the applicable wavelength range, by employing the dual-wavelength injection-seeded oscillation with a Ti: sapphire laser, as an example.

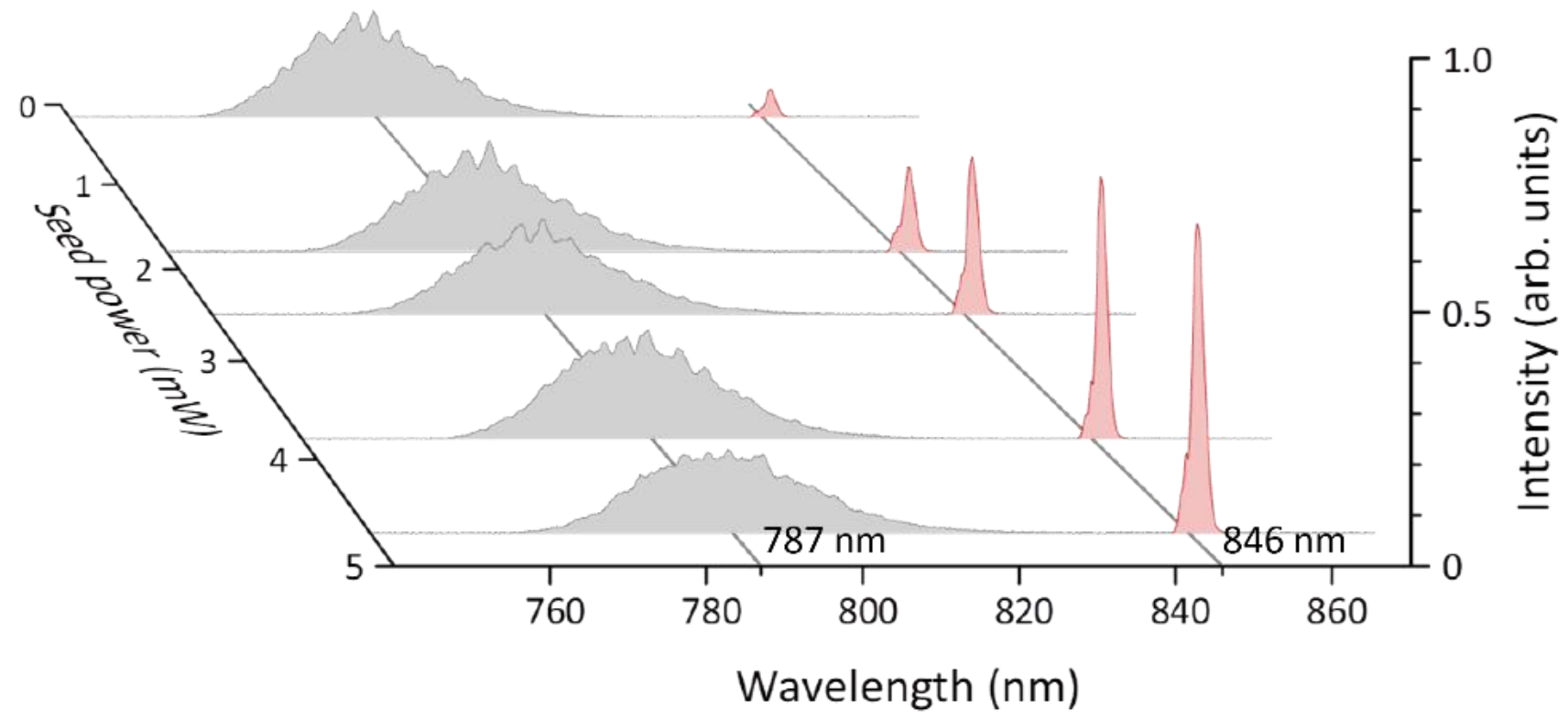

**Fig. 2. Test of suppressing the free-running oscillation by increasing injected power of the seed laser at a wavelength (846 nm) far-off the gain center.**

Among the possible dual-wavelength combinations, the most stringent requirements are placed on the simultaneous oscillation with combination of the gain-center and -tail wavelengths (e.g., the combination of $\lambda_1$ and $\lambda_3$ in Fig. 1). This also remains as it is, in the case of multi-wavelength oscillation. Below, we will discuss the methodology by focusing on this case.

There are two possible mechanisms for extending the applicable gain-edge wavelength: increasing the seed laser power at the gain-edge wavelength, and reducing the gain at the gain center to mitigate the gain imbalance between the gain center and edge.

Figure 2 presents a test according to the former approach in mind. We injected a seed laser radiation at a wavelength in the gain tail ($\lambda_3$: 846 nm) into a Ti:sapphire laser while increasing the seed-laser power. Around the gain-center wavelength ($\lambda_1$: 787 nm), we found the broad free-running oscillation spectrum that was bulit up from spontaneous emissions (the gray spectrum in Fig. 2). While the free-running oscillation was slightly reduced as the seed power was increased at $\lambda_3$, the free-running oscillation spectrum almost remained unsuppressed even when the seed power was raised to 4.7 mW. In contrast, at the gain-center wavelength, $\lambda_1$, the free-running oscillation was completely suppressed even with the seed power below 100 µW. Namely, the above result implies that the simultaneous oscillations at the gain-center and -tail wavelengths never be achieved, even if we had provided a seed power ratio of 1 to 40 between the gain-cneter and the gain-tail wavelengths. The former approach is impracticai.

Next, we consider an alternative approach that mitigates the gain imbalance between the gain-center and -tail wavelengths. This approach may involve two methods: inserting wavelength-dependent optical elements inside the laser resonator [14] or introducing wavelength-dependent reflectivity on the output-coupling mirror. While the former is commonly employed, inserting such optical elements in the laser resonator, inevitably introduces losses that significantly reduces the amount of output power. In contrast, in the latter method, introducing a wavelength-dependent reflectivity on the output-coupling mirror, supression of the reflectivity never induce power loss, but can be readily converted to the output power.

Given these considerations, we select the latter approach of implementing wavelength-dependent reflectivity on the output-coupling mirror. The optimal reflectivity characteristics for this purpose would be qualitatively expected to suppress the reflectivity at the gain-center wavelength while maintaining a high reflectivity at the gain tail (Figure 1, dual-OC). The key question then becomes how to quantitatively determine these reflectivity characteristics. Here, we present this methodology and furthermore demonstrate, using an actual output-coupling mirror fabricated based on this approach, that the applicable wavelength-range for dual-wavelength oscillation is indeed expanded as designed.

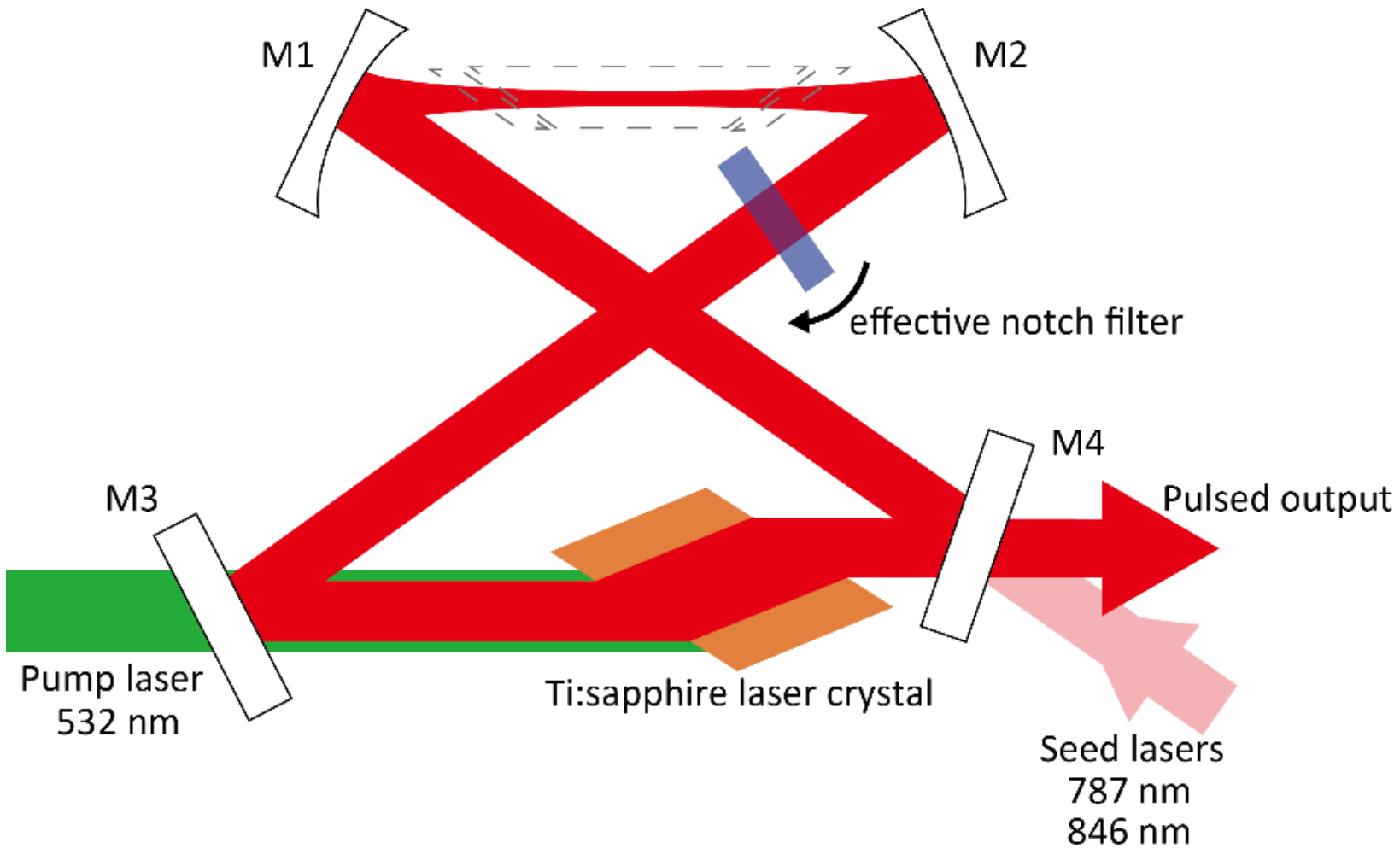

**Fig. 3. Schematic layout for experimentally estimating the wavelength-dependent loss required for extending the available wavelength range of a dual-wavelength injection-seeded Ti: sapphire laser.** M1, M2, concave (radius, 100 mm) and M3, plane, total-reflection mirrors with a reflectivity of > 99% over 750 - 850 nm. M4, output coupler with a standard property of a constant reflectivity (**66.5%**) over 750 - 850 nm. Blue plate, the optical device employed as an effective notch-type filter, mounted on a rotary stage.

As the first step in designing the reflectivity characteristics of the output-coupling mirror, we estimated a required gain-attenuation rate around the gain center ($\lambda_1$: 787 nm) to suppress the free-running oscillation while enabling injection-seeded operation with a single wavelength ($\lambda_3$: 846 nm) around the gain tail.

As shown in Fig. 3, we inserted an optical element functioning as an effective notch filter (ENF: standard total-reflection mirror at 633nm, at an incidence of angle, 45°) in the Bow-tie ring Ti:sapphire laser-resonator. This element has a transmission profile that introduces loss at the gain center while having minimal effect around the gain-tail. By gradually adjusting the insertion angle of this optical element from normal incidence, we controlled the attenuation rate to be added to the gain center.

In this configuration, the Bow-tie ring Ti:sapphire laser consisted of three total-reflection mirrors (reflectivity > 99%; M1, M2: concave mirror, radius of curvature of 100 mm, M3: plane mirror) and one output-coupling mirror, M4 (reflectivity: 66.5 ± 0.5%). Either of the optical elements had standard flat-reflectivity properties across 750 - 850 nm. The beam diameter inside the ring resonator was designed to be 1.2 mm at $1/e^2$ at the Ti: sapphire crystal. (more details of the laser-resonator will be reported elsewhere)

***a***

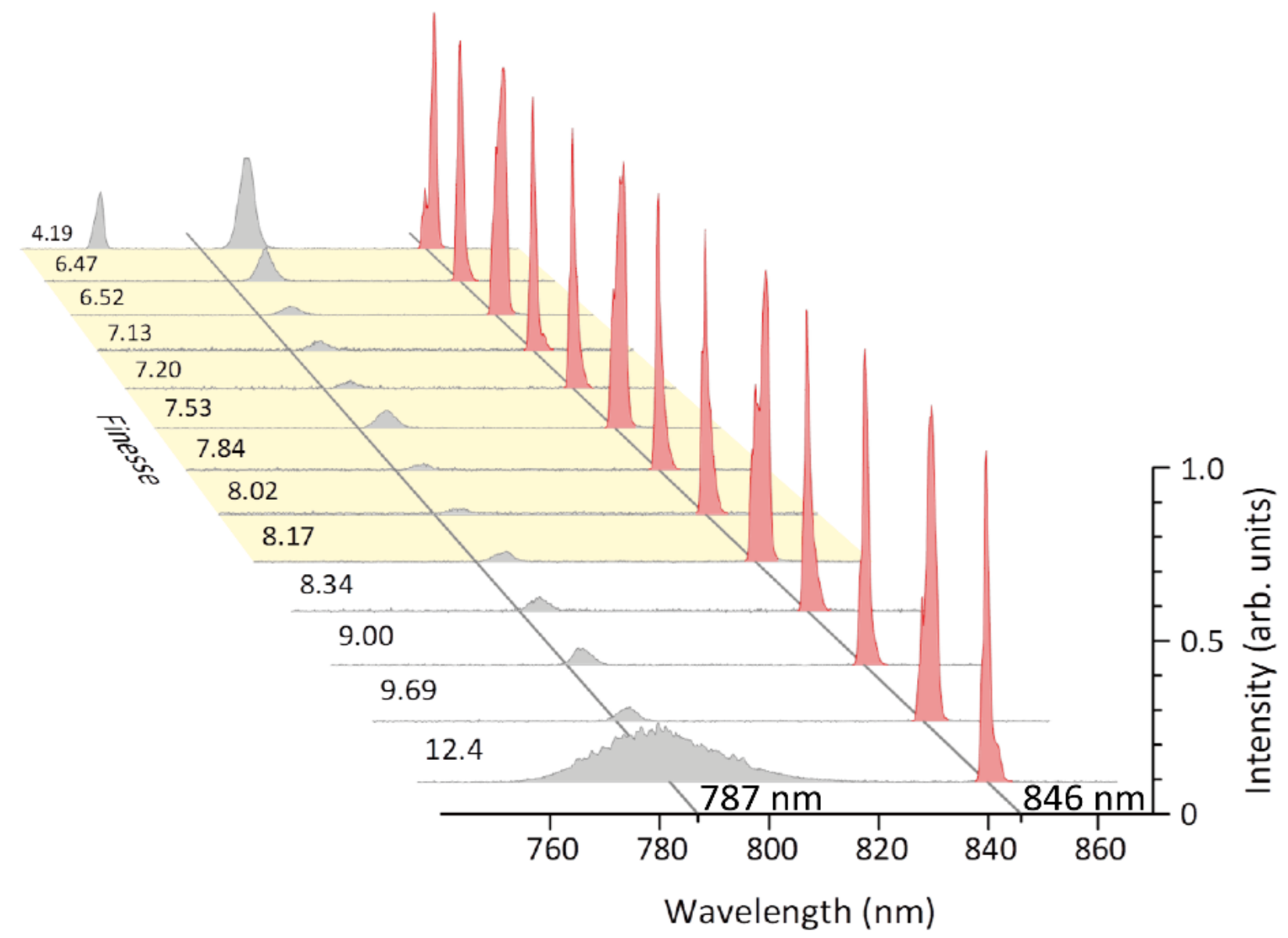

***b***

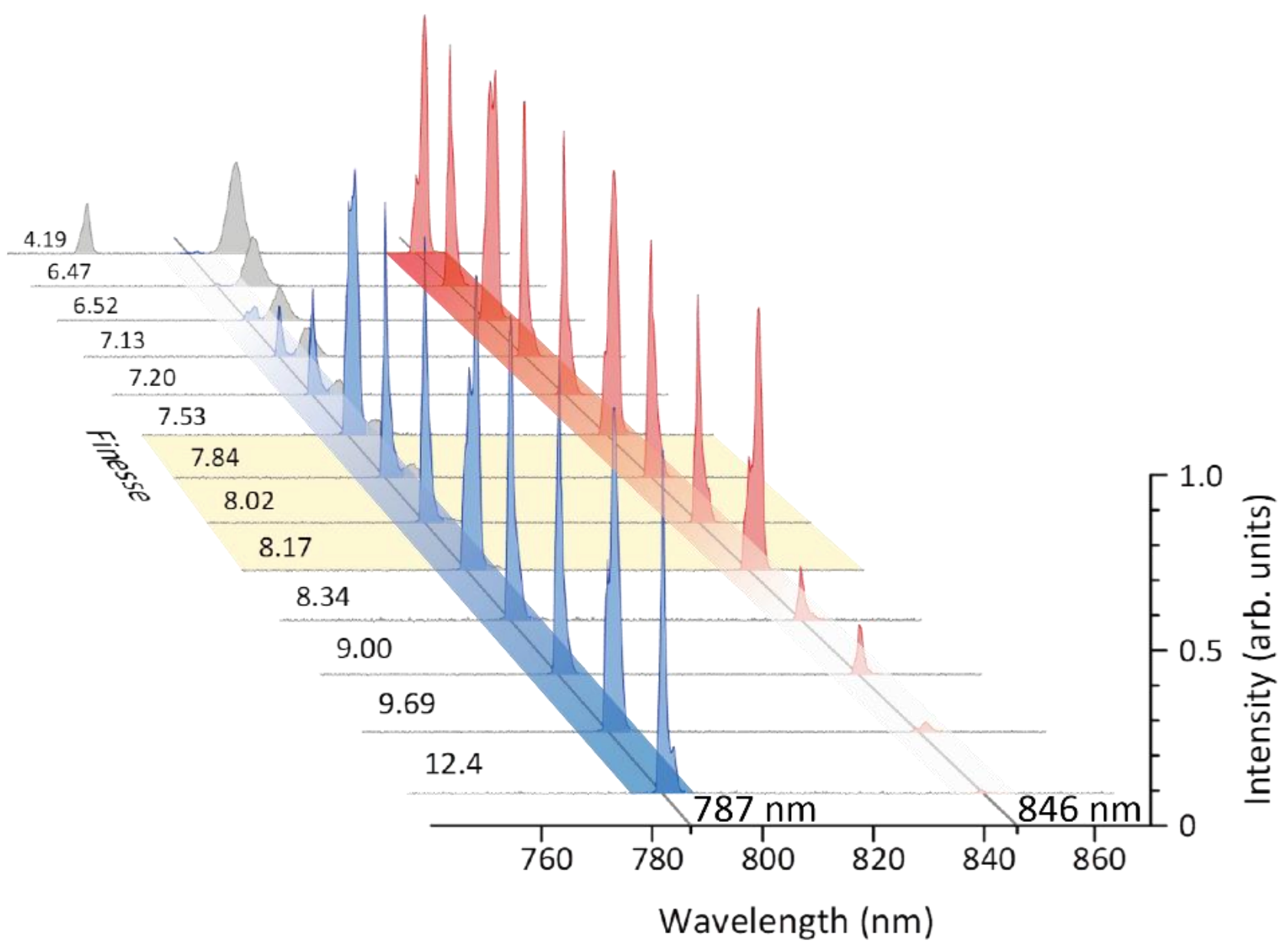

**Fig. 4. Estimation of the wavelength-dependent loss required for extending the available wavelength range of a dual-wavelength injection-seeded Ti: sapphire laser.** Tests of injection-seeded pulsed oscillation, ***a***, with a single seed-laser injected at a wavelength far-off the gain center, 846 nm, and ***b***, with a dual-wavelength seed lasers injected at wavelengths around and far-off the gain center, 787 nm and 846 nm, respectively. Notations of the finesse are measured at 787 nm.

In practice, the losses added to the gain-center wavelength, $\lambda_1$ (787 nm), were manipulated between 7 - 84%. To quantify these losses in terms of total resonator-losses, which are directly related to the laser operation, we precisely measured the resonator finesses using the 787-nm seed laser at each of the installation angles. The above-mentioned additional losses corresponded to a resonator finesse ranging from 4.19 to 12.4, which included all losses with the other resonator components.

Figure 4 shows the results when the seed-laser radiation was introduced at $\lambda_3$ (846 nm) to lead a single-wavelength injection-seeded oscillation in the gain-tail wavelength. Here, the pump laser at 532 nm was set to an energy of 56 mJ (repetition rate, 10 Hz), while the seed laser power at $\lambda_3$ in the gain tail was adjusted to 3.2 mW. The pulsed output spectrum was measured using a compact spectrometer (USB4000-UV-VIS, Ocean Optics Co.).

As seen in Fig. 4, when the losses exceeding a finesse of 8.17 (with an additional loss of 33%) were added to the gain-center wavelength, the free-running oscillation originating from spontaneous emission at the gain-center wavelength was almost suppressed, and the pulsed output was drawn in the seed laser radiation injected in the gain tail.

It should be noted that, under conditions with additional losses corresponding to a finesse below 8.17, the residual free-running oscillation spectra were found on the longer wavelength side of the gain center, because the losses by ENF were given in the limited wavelength-widths around the gain center (depending on the insertion angle). Since the objective of this evaluation was to determine the additional loss that can suppress the free-running oscillation at the gain center (787 nm) while enabling the single-wavelength injection-seeded oscillation in the gain tail (846 nm), the residual free-running oscillation spectra remained on the longer wavelength side did not affect this evaluation test. (for more detailed information, see the supplementary material.)

Having estimated the requirement for suppressing the free-running oscillation that builds up from spontaneous emissions around the gain center, we then introduced the seed laser radiation at 787 nm (gain center) in each of the configurations shown in Fig. 4a to examine dual-wavelength, injection-seeded oscillations at the gain-center and -tail wavelengths (specifically, determining the requirements for achieving a power ratio of the injection-seeded pulsed outputs with 1 to 1).

Here, we regarded the ratio of 1 to 15 at maximum between seed power at the gain center and that in the gain tail to be practical, although the ideal ratio is 1 to 1 in terms of controllability of the pulsed output-power ratios [7].

As expected by the results obtained in Fig. 4a, the loss added to the gain center proved insufficient when the finesse at the gain center was larger than 8.34, thereby, the injection-seeded oscillation in the gain tail could not be built up by being overshadowed by that at the gain center, even though we provided a seed-power ratio of up to the maximum between the gain center and gain tail. This demonstrates that suppressing free-running oscillation is a prerequisite.

Meanwhile, when the loss was further added at the gain center to attain a finesse of 8.17—sufficiently suppressing the free-running oscillation at the gain center—as shown in Fig. 4b, injection-seeded pulsed oscillations were achieved both at the gain center and in the gain tail with a power ratio of nearly 1 to 1. Such dual-wavelength injection-seeded oscillations were sustained until the loss corresponding to a finesse of 7.53 was provided.

However, further increasing losses, with the finesses of 7.13–7.20, beyond this point resulted in excessive losses, causing the injection-seeded oscillation at the gain center to be overpowered by that in the gain tail. Additionally, when losses equivalent to or greater than those corresponding to a finesse of 6.47 were introduced, the injection-seeded oscillation at the gain center was no more observed.

From these results, we concluded that the appropriate range of losses to be applied to the gain center to achieve comparable injection-locked pulsed-output powers both at the gain center and in the gain tail should be 7.53–8.17 in finesse. In this evaluation test, the finesse at 846 nm in the gain tail was maintained within the range of 11.0–11.5.

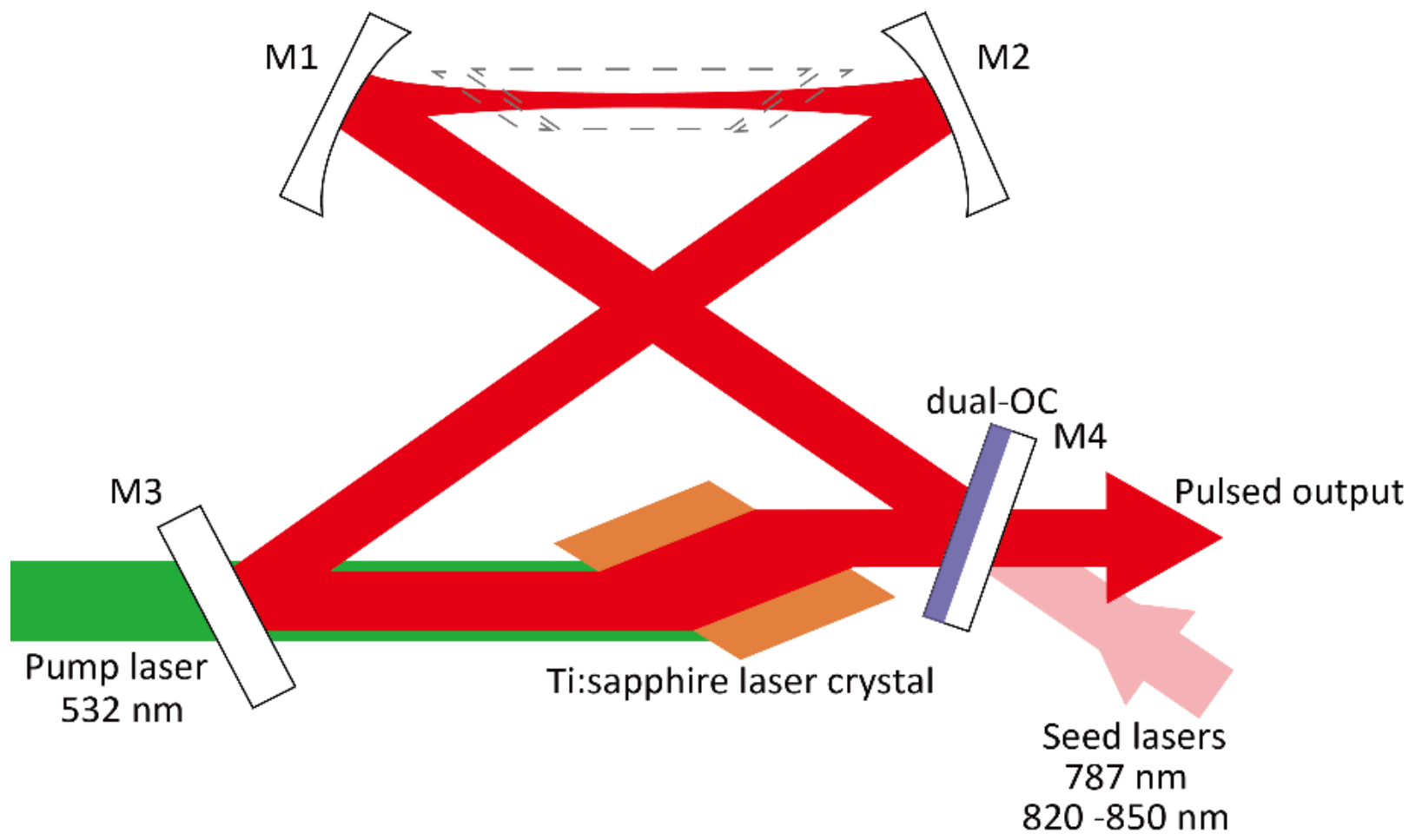

**Fig. 5. Schematic illustration of the dual-wavelength injection-seeded Ti:sapphire laser with the dual-OC (M4).** Total reflection mirrors (M1, M2, concave with a radius of 100 mm, and M3, plane) with a reflectivity of > 99% over 750 - 850 nm.

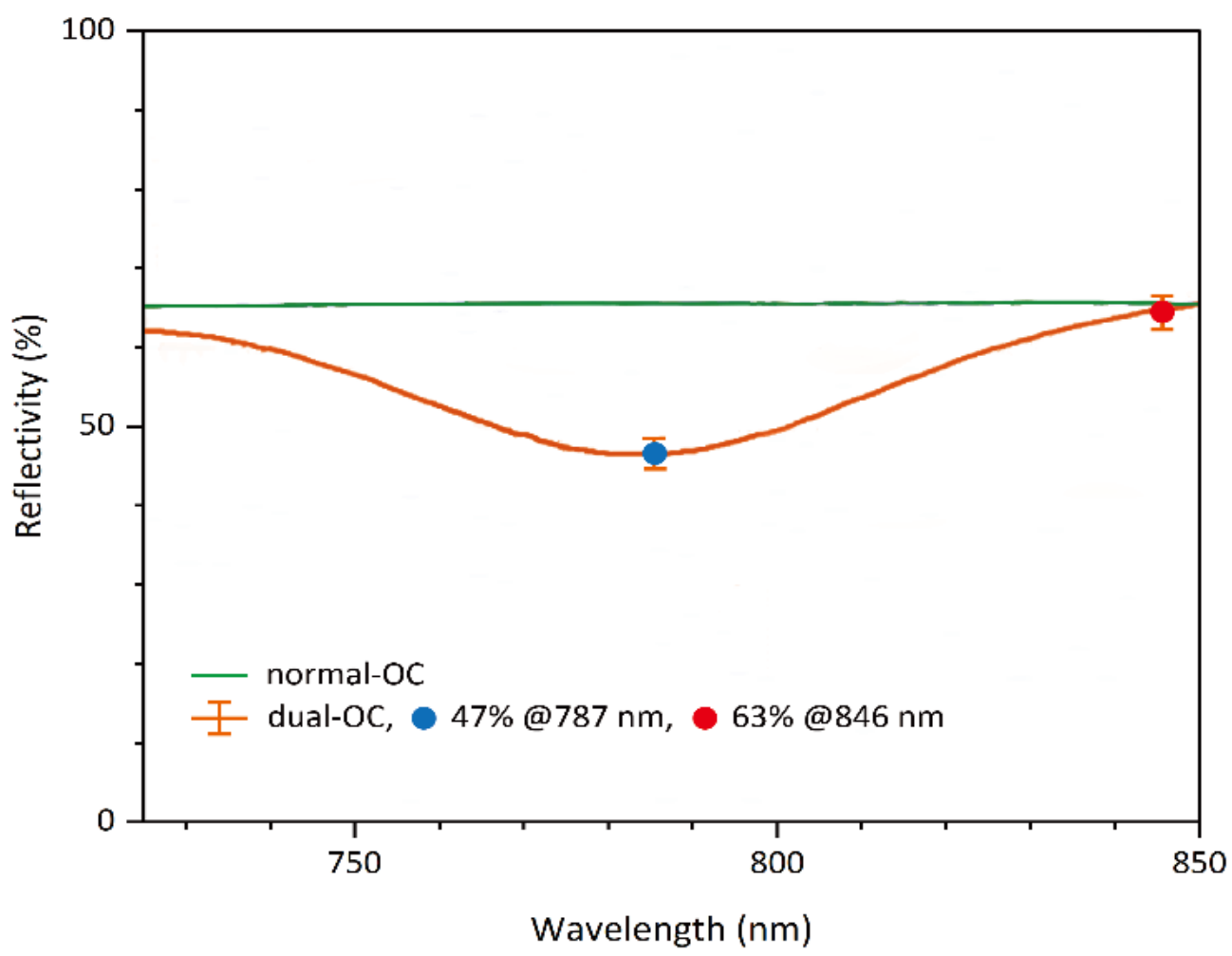

**Fig. 6. Designed and measured reflectivities of the dual-OC.** Orange curve with bars, designed reflectivity curve with an allowable range of +2.5%. Solid circles, measured reflectivities of the dual-OC, around (blue) and far-off (red) the gain center. Green curve, reflectivity of the normal-OC is shown for comparison.

In the evaluation experiments using ENF, we determined the appropriate finesses for expanding the available wavelength range of the dual-wavelength injection-seeded oscillations to be 7.53-8.17 for 787 nm while maintaining the finess at 846 nm in the range of 11.0–11.42. These values were then converted into reflectivity of the output-coupling-mirror, R, by using Eq. 1. Here, coefficient, A, represents the inherent total losses present inside the laser cavity (reflection losses of the crystal surfaces and mirrors)— 8.2% at 787 nm, 8.0% at 846 nm.

$$Finesse = \frac{\pi^4 \sqrt{R}}{1 - \sqrt{RA}} \tag{1}$$

The derived reflectivities were 46.7–49.9% at 787 nm and 60.8–62.1% at 846 nm. The tolerance required to be included in the manufacturing process was $\pm$2.5% at largest. Since suppressing the free-running oscillation at the gain center was first priority, we specified the upper limit of the reflectivity at 787 nm as 49.9% including the tolerance of $\pm$2.5%, that is, requested the specification of reflectivity as 47.4 ± 2.5% for 787 nm. Similarly, we specified the specification of reflectivity as 63.3 ± 2.5% for 846 nm by regarding the lower limit of reflectivity as 60.8%. Regarding the reflectivities between the gain center and tail, we requested a smooth reflectivity-variation with a structure of convex down to convex up as the wavelengths were longer (Figure 6, dual-OC configuration). The reflectivities of the fabricated output-coupling mirror, in practice, were measured to be 47% at 787 nm and 63% at 846 nm, confirming within the designed reflectivity range.

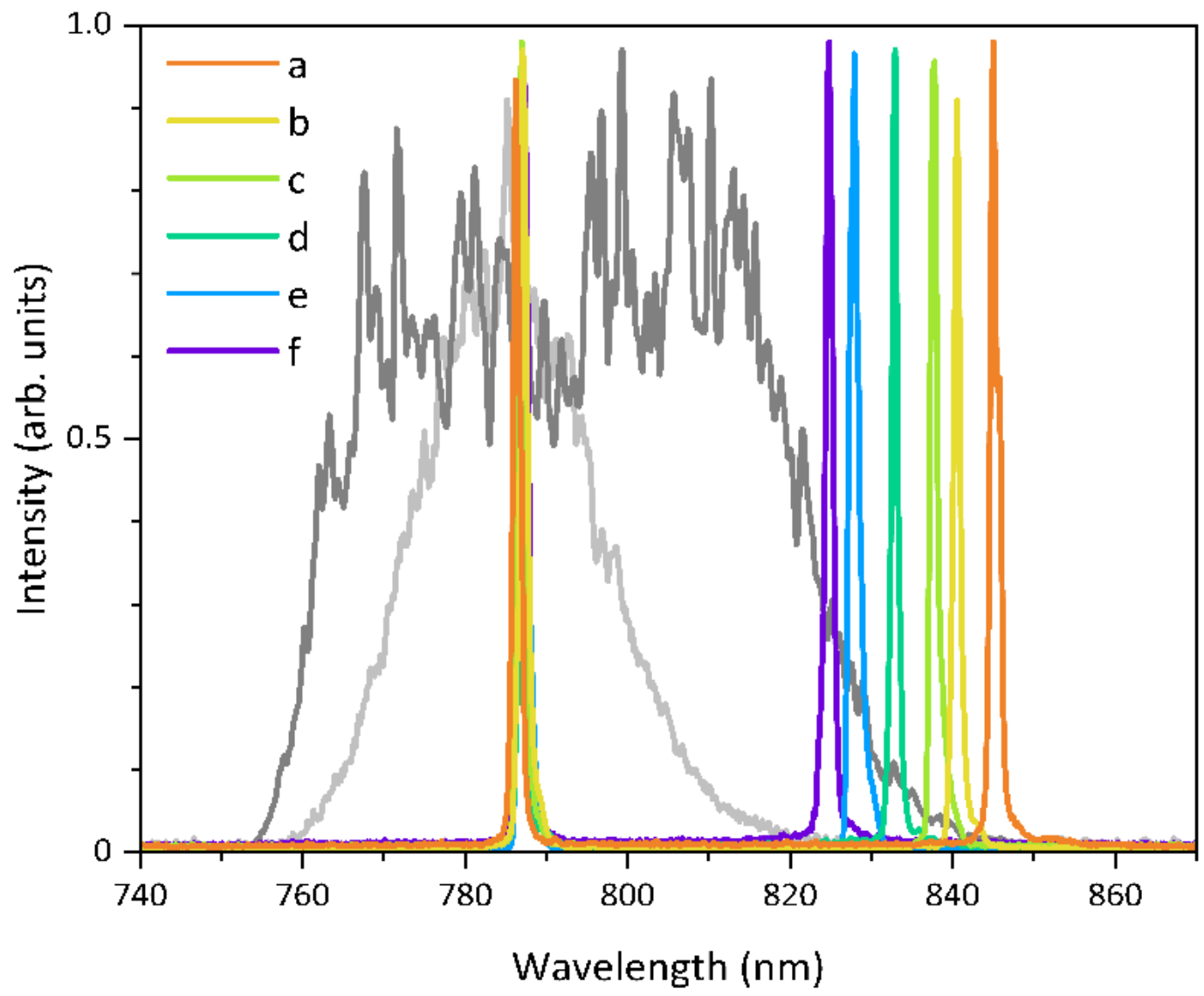

**Fig. 7. Pulsed output spectra of the injection-seeded Ti:sapphire laser with a variety combinations of dual wavelengths.** a: (786.9537 nm, 824.3666 nm), b: (786.9545 nm, 827.8072 nm), c: (786.9534 nm, 832.7388 nm), d: (786.9537 nm, 837.9514 nm), e: (786.9536 nm, 840.3978 nm), f: (786.9530 nm, 845.5836 nm). Spectra of free-running oscillations are also shown; light gray spectrum, operated by employing a standard output coupler (normal-OC), dark gray spectrum, the designed output coupler (dual-OC).

We integrated the fabricated output-coupling mirror (dual-OC) into the laser cavity, as shown in Fig. 5 and conducted tests for dual-wavelength oscillations both at the gain center and in the gain tail. The pump energy of a 532-nm laser was set to 57 mJ with a beam diameter of 1.2 mm at $1/e^2$ at the Ti: sapphire crystal surface, where the pulse duration and the repetition rate were 10 ns and 10 Hz, respectively.

The dark gray line in Fig. 7 represents a spectrum of the observed free-running oscillation without introducing the seed-laser radiations. The reduction of finesse around the gain center wavelength resulted in a less pronounced gain-narrowing behavior compared to the case using a normal-OC (the light gray lines in Figs. 1 and 7), demonstrating a broader oscillation spectrum.

Here, we injected seed laser radiations at both the gain-center wavelength (786.9530 nm) and gain-tail wavelength (845.5836 nm) for inducing a dual-wavelength injection-seeded oscillation. By adjusting the seed-laser power ratio to achieve a nearly equal pulsed-output powers, we obtained those at a seed power ratio of 1 to 7 between the gain center and tail (f, orange spectrum in Fig. 7).

We then tested other dual-wavelength injection-seeded oscillations with a variety of combinations of the gain-center wavelength (787 nm) and gain-tail wavelengths (824 – 840 nm, a – e in Fig. 7). For all combinations, nearly-equal pulsed outputs could be achieved by properly adjusting the seed power ratio between 2 to 1 – 1 to 7.

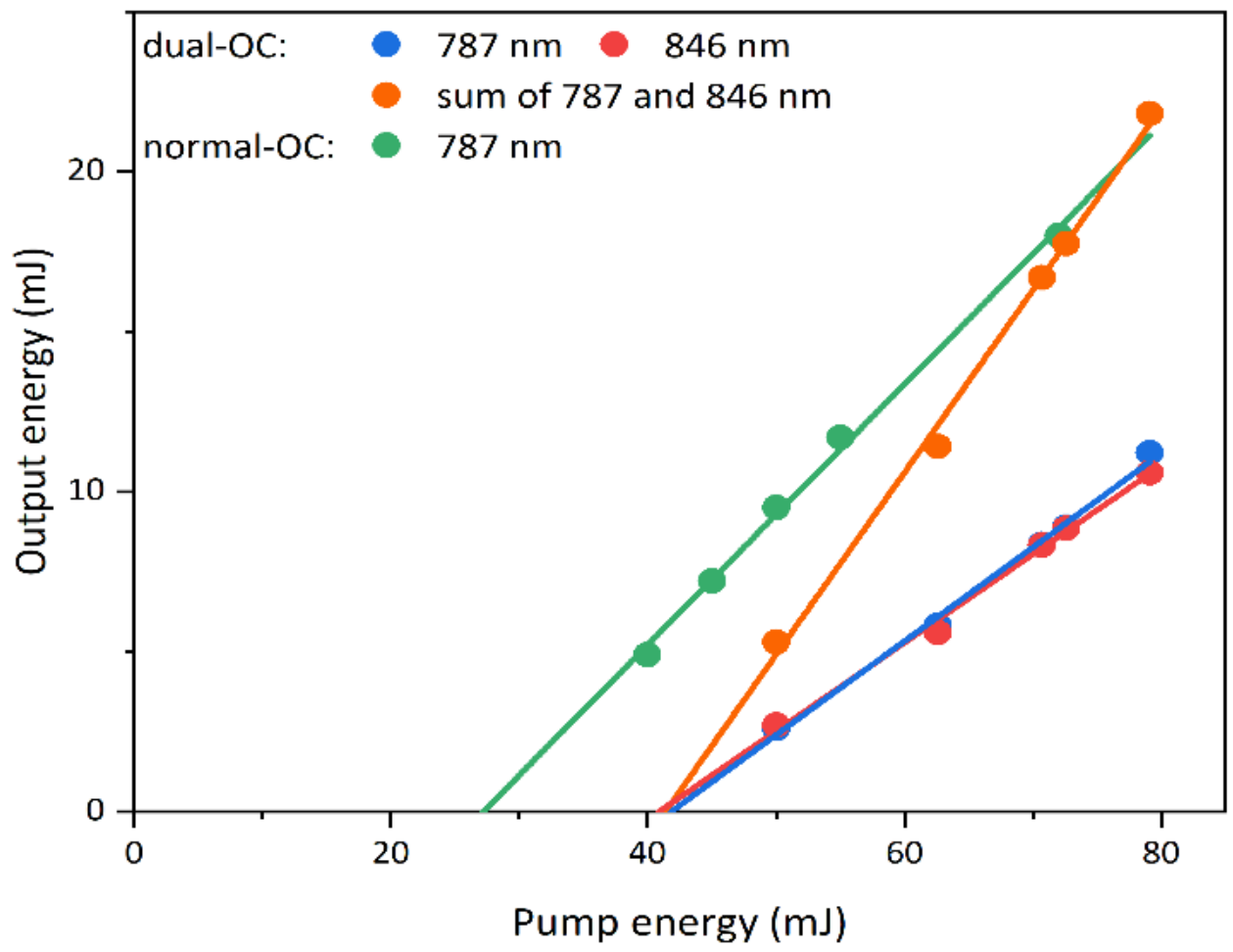

**Fig. 8. Pulsed output properties of the dual-wavelength injection-seeded Ti: sapphire laser operated with the dual-OC, at dual seed-wavelengths injected around the gain center (blue solid circles) and far-off the center (red solid circles)**, where orange solid circles indicate sum of output powers of those of blue and red solid-circles. For comparison, output property with the normal-OC (green solid circles) is also shown, where the Ti: sapphire laser is operated with a single-wavelength seed laser around the gain center.

Finally, we present the output-power characteristics for the dual-wavelength injection-seeded oscillation, corresponding to the case f in Fig. 7. Measurements were done under the condition where nearly-equal output powers were achieved in the gain center and gain tail. The output-power characteristics exhibited nearly identical in either of the gain center and gain tail, regarding oscillation thresholds, slope efficiencies, maximum output-powers, and maximum energy-efficiencies, which were 42 mJ, 29%, 11.2 mJ/14.2%; and 41 mJ, 28%, 10.6 mJ/13.4% for the injection-seeded oscillations at 787 nm (blue solid circles in Fig. 8) and 846 nm (red solid circles in Fig. 8), respectively, where the maximum pump energy was set to 79 mJ.

In the total output-power characteristics obtained by taking sum of those at 787 nm and 846 nm, they amount 41 mJ, 57%, 21.8 mJ/27.6% (orange solid circles in Fig. 8). For comparison, we also list the output characteristics for the single-wavelength injection-seeded oscillation at the gain-center, 787 nm (green solid circles in Fig. 8), using a cavity configured with normal output-coupling mirror. The oscillation threshold, slope efficiency, maximum output-power, and maximum efficiency were 27 mJ, 41%, 18.0 mJ/25.0%, respectively, where the maximum pump energy was set to 72 mJ.

As the designed output-coupling mirror (dual-OC) inherently requires lower output-coupling reflectivity, the oscillation threshold in the dual-wavelength oscillation increased compared to that in the normal-OC configuration (27 -> 41 mJ). However, in turn, the slope efficiency improved significantly (41% -> 57%), thus, the output power remained nearly unchanged when sufficient pump energy was provided (> 70 mJ), as exhibited in Fig. 8. The design of wavelength-dependent cavity property was achieved not by conventional methods placing wavelength-dependent components inside the laser cavity, which introduced significant losses, but rather by designing wavelength-dependent property into the reflectivity of output-coupling mirror. This approach enabled us to achieve the purpose, extension of the available wavelength-range of the dual-wavelength injection-seeded oscillations, without introducing any losses in the cavity.

**Conclusion** We have described the methodology that expands an applicable wavelength range in the dual-wavelength injection-locked laser. We have shown that it can be achieved without introducing losses in the resonator, by properly designing a wavelength-dependent characteristics of the output-coupling mirror based on an experimental measurement. Using the designed output-coupling mirror, in practices, we have demonstrated dual-wavelength oscillations with an equal output power, across a variety of wavelength combinations of near the gain center and around the gain tail, in which the wavelength separation amounts 60 nm at the most.

Injection-seeded nanosecond pulsed lasers can simultaneously realize both high power (> MW) and high spectral-purity (< tens of MHz), making them particularly useful for studying nonlinear optical phenomena, involving those with isolated atomic/molecular systems, and also studying atmospheric physics. The laser technology presented here, expansion of the applicable wavelength range in the dual-wavelength injection-seeded lasers, may open diverse applications, including simultaneous measurement of Ca (422.7918 nm × 2 = 845.5836 nm) and $Ca^+$ (393.4770 nm × 2 = 786.9540 nm) layers originating from meteoroids, appearing in the upper atmosphere (altitude: >80 km) [15].

**Funding.** This work was supported by Grant-in-Aid for Scientific Research (S) No. 20H05642. We also acknowledge financial support from Nichia Corp.

**Acknowledgments.** The authors thank S. Miyoshi for supporting the experiments for this work. They also thank T. Nakamura and M. Ejiri for useful discussions.

**Disclosures.** The authors declare no conflicts of interest.

**Data availability.** Data underlying the results presented in this paper are not publicly available at this time but may be obtained from the authors upon reasonable request.

**Supplementary**

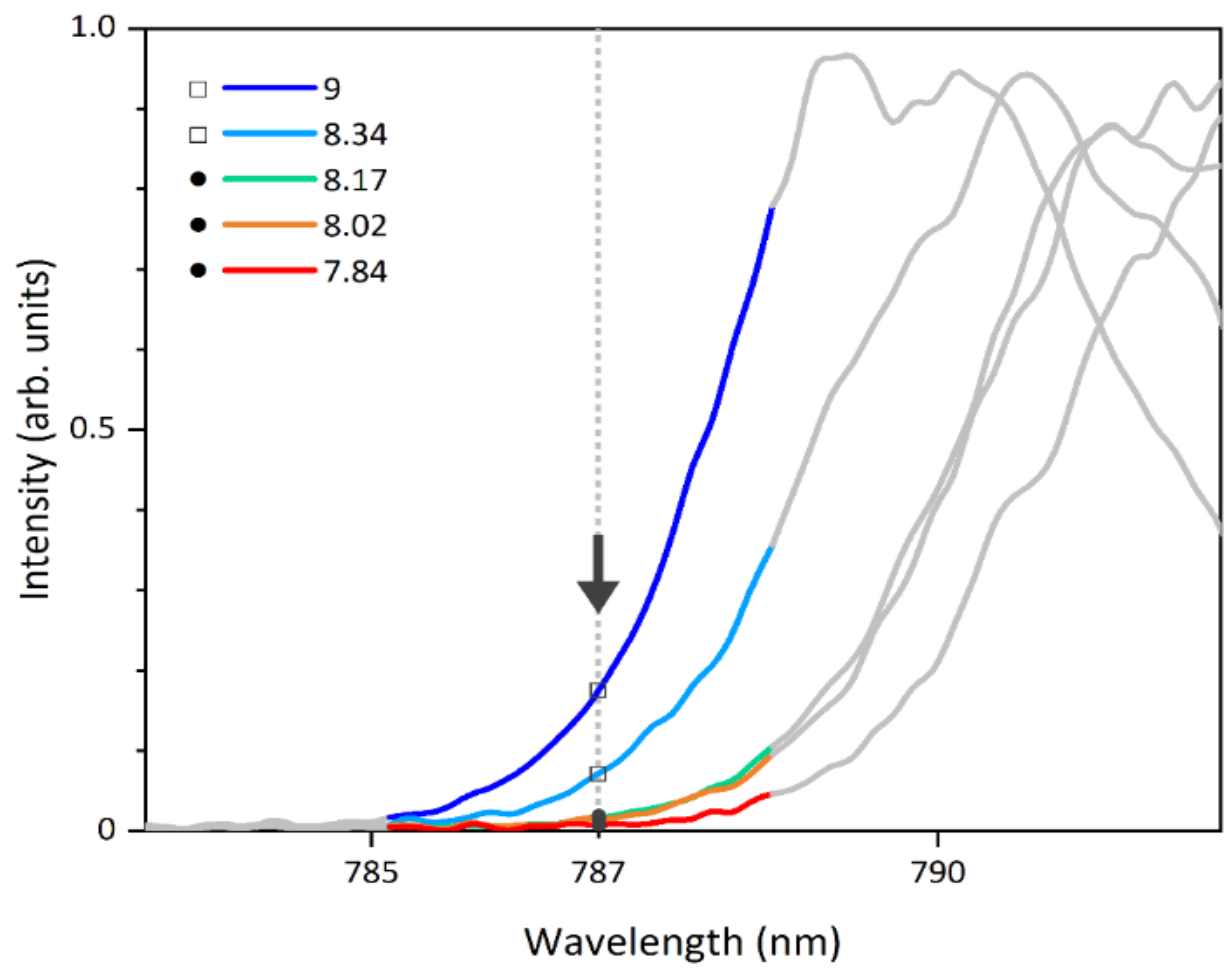

**Fig. S1. Detail structures of Fig. 4a.** Suppressions of free-running oscillations at the gain center, 787 nm, are seen in detail as the given losses are increased. Due to a property of the optical device employed as a notch filter, the free-running oscillation profiles are pushed away to the longer wavelength ranges as the losses are increased by changing incident angles of the optical device. The free-running oscillation is sufficiently suppressed when the loss is given with an amount larger than that corresponding to a finesse of 8.17 at the gain center, 787 nm.